\documentclass[conference]{IEEEtran}
\usepackage{amsfonts}
\usepackage{amssymb}
\usepackage[cmex10]{amsmath}
\usepackage{graphicx}
\usepackage{subfigure}
\usepackage{verbatim}
\usepackage{cite}
\usepackage{booktabs}
\usepackage[ruled]{algorithm2e}
\usepackage{multirow}
\usepackage{diagbox}
\usepackage{slashbox}
\usepackage{bm}
\usepackage{amsthm}
\usepackage{mathrsfs}

\newcommand{\bX}{{\mathbf X}}
\newcommand{\bx}{{\mathbf x}}

\newcommand{\bPsi}{\Psi}
\newcommand{\bpsi}{{\psi}}

\newcommand{\bpi}{{\boldsymbol\pi}}

\newtheorem{Def}{Definition}
\newtheorem{Pro}{Proposition}

\newtheorem{Lem}{Lemma}
\allowdisplaybreaks
\newcommand{\tabincell}[2]{\begin{tabular}{@{}#1@{}}#2\end{tabular}}

\ifCLASSINFOpdf
\else
\fi
\IEEEoverridecommandlockouts
\begin{document}

\title{Principled Random Finite Set Approximations of Labeled  Multi-Object Densities}
%

\author{
Suqi Li, Wei Yi, Bailu Wang and Lingjiang Kong\\
\IEEEauthorblockA{{University of Electronic Science and Technology of China, School of Electronic Engineering, Chengdu City, China} \\
{Email: kussoyi@gmail.com}}}
\maketitle

\begin{abstract}
As a fundamental piece of multi-object Bayesian inference, multi-object density has the ability to describe the uncertainty of the number and values of objects, as well as the statistical correlation between objects, thus perfectly matches the behavior of multi-object system. However, it also makes the set integral suffer from the curse of dimensionality and the inherently combinatorial nature of the problem. In this paper, we study the approximations for the universal labeled multi-object (LMO) density and derive several principled approximations including labeled multi-Bernoulli, labeled Poisson and labeled independent identically clustering process based approximations. Also, a detailed analysis on the characteristics (e.g., approximation error and computational complexity) of the proposed approximations is provided. Then some practical suggestions are made for the applications of these approximations based on the preceding analysis and discussion. Finally, an numerical example is given to support our study.
\end{abstract}

\IEEEpeerreviewmaketitle

\section{Introduction}

The object of multi-object inference is the simultaneous estimations of   the number and individual states of objects, with the applications spanning many areas, such as forestry \cite{refr:forest}, biology \cite{refr:biology}, physics \cite{refr:physics}, computer vision \cite{refr:computer-vision}, wireless
networks \cite{refr:network}, communications \cite{refr:communication}, multi-target tracking \cite{refr:tracking-1,refr:tracking-2}, and robotic\cite{refr:robotics}. The states of objects are essentially a type of point pattern whose probability model is the point process (specifically simple finite point processes or random finite sets (RFS)) derived from stochastic geometry.  To handle the statistics of point process, the point process theory \cite{refr:point_process} provides  powerful mathematic tools. Furthermore, finite set statistics (FISST) \cite{refr:Mahler_book} pioneered by Mahler is also proposed to  deal with RFSs based on a notion of integration and density that is consistent with point process theory.

In point process theory or FISST, \textit{multi-object  probability density} is a fundamental descriptor  for point processes, for it is central to the foundation of multi-object estimation:  multi-object Bayesian inference. Multi-object densities have  ability to describe the uncertainty of the number and values of objects, as well as the statistical correlation between objects, which perfectly match the behavior of multi-object system.  Also for this reason the multi-object probability density usually has  a complicate structure, and thus the computation of the multi-object posterior via Bayes rule is usually intractable  with the set integral suffering from the curse of dimensionality and the inherently combinatorial nature of the problem.

To solve these problems, several tractable approximations of multi-object  density, such as Poisson, i.i.d clustering processes and multi-Bernoulli density \cite{refr:Mahler_book} have been proposed.  These approximations  dramatically simplify the set integral and  lead to several efficient  multi-object filtering algorithms, i.e., probability hypothesis density (PHD) filter \cite{refr:PHD}, cardinalized PHD filter \cite{refr:CPHD} and multi-Bernoulli filter \cite{refr:MeMber_filter1,refr:MeMber_filter}.

 Recently, with the development of FISST, the research focus turns to the labeled multi-object (LMO) density  and the corresponding labeled set filters \cite{refr:label_1,refr:label_2,refr:label_3,refr:label_4,refr:label_5,refr:label_6,refr:label_7} due to their advantageous performance compared to previous (unlabeled) RFS, such as identifying object identities, simplifying the multi-object transition kernel.   In \cite{refr:label_1,refr:label_2}, Vo \textit{et al}. proposed  a class of generalized labeled multi-Bernoulli (GLMB) density \footnote{GLMB distribution is also simply named as Vo-Vo distribution by Malher in his book \cite{refr:tracking-2} first time. } which is  a conjugate prior and also closed under the Chapman-Kolmogorov equation for the standard multi-object system in Bayesian inference.  Later, in order to further improve the computation efficiency, principled approximations of GLMB density are proposed, including labeled multi-Bernoulli (LMB) density \cite{refr:label_5} and marginalized $\delta$-GLMB (M$\delta$-GLMB) density \cite{refr:label_3}.  These two approximations are computationally cheaper, as well as reserve the key statistical properties of the full multi-object density and thus have decent accuracy.

The GLMB density is an accurate representation of the prior and posterior LMO density  in the standard multi-object system, but is not necessarily for the universal LMO density. Typical applications of the universal LMO density include, for instance,  the multi-object estimation with non-separable multi-object likelihood \cite{closely-spaced-1,closely-spaced-2,closely-spaced-3,closely-spaced-4,closely-spaced-5,closely-spaced-6}.     Papi \textit{et al}. proposed a $\delta$-GLMB density approximation  of the universal LMO density with the cardinality distribution and the first moment preserved, as well as the Kullback-Leibler divergence (KLD) minimized in \cite{refr:label_6}.  The high accuracy of the $\delta$-GLMB density approximation is verified by the $\delta$-GLMB filtering algorithm proposed in \cite{refr:label_6}, but note that the set integral of the $\delta$-GLMB density still suffers from the combinational nature of problem.


Inspired by the work of \cite{refr:label_6}, we study the principled approximations for the universal LMO density with other label RFS,  and our contributions are two-fold:

Firstly, we propose a series of principled approximations preserving the key statistical characters of the original LMO density and improving the computational efficiency of set integral, namely,  LMB density, labeled Poisson (LP) process and labeled independent identically (LIID) clustering process approximations. Specifically,  the LMB density approximation preserves the labeled first order moment and minimize the KLD from the original LMO density. The LP process approximation only preserves the unlabeled first order moment, while  LIID clustering process approximation  preserves both the unlabeled first order moment and the cardinality distribution of original LMO density.

Secondly, to provide a guidance for the practical application, we compare the  relative merits  of  the aforementioned approximations and $\delta$-GLMB density approximation systematically in terms of approximation error and computational complexity, and give some practicable suggestions for the selection of these approximations based on the solid theoretical analyses.  An numerical example is provided to support our analysis on different approximations.
\section{Background}
\subsection{Notation}
We adhere to the convention that single-object states are
represented by lowercase letters, e.g., $\bx$, while multi-object
states are represented by uppercase letters, e.g., $\bX$, $X$.  To distinguish labeled states and distributions from the
unlabeled ones, bold-type letters are adopted for the labeled
ones, e.g., $\bx$, $\bX$, $\bpi$. Moreover, blackboard bold letters represent spaces, e.g., the
state space is represented by $\mathbb{X}$, the label space by $\mathbb{L}$. The collection of all finite sets of $\mathbb{X}$
is denoted by $\mathcal{F}(\mathbb{X})$, and the collection of all finite sets  of $\mathbb{X}$ with cardinality $n$ is denoted by $\mathcal{F}_n(\mathbb{X})$.

%

We use the multi-object exponential notation
\begin{equation}\label{multi-object exponential notation }
  h^{X}\triangleq\prod_{x\in X}h(x)
\end{equation}
for real-valued function $h$, with $h^\emptyset=1$ by convention.
To admit arbitrary arguments like sets, vectors and integers, the generalized Kronecker delta function is given by
\begin{equation}\label{delta}
  \delta_Y(X)\triangleq\left\{\begin{array}{l}
\!\!1, \mbox{if}\,\,\,X=Y\\
\!\!0, \mbox{otherwise}.
\end{array}\right.
\end{equation}
The inclusion function $1_Y(X)$ is given by
\begin{equation}\label{eq:A2}
  1_Y(X)=\left\{ \begin{array}{ll}
\ 1, & \mbox{if}\,\,X\subseteq Y\\
\ 0, & \mbox{if otherwise}.\\
\end{array} \right.
\end{equation}
If $X$ is a singleton, i.e., $X=\{x\}$, the notation $1_Y(\{x\})$ is used instead of $1_Y{\{x\}}$.

 The integrals  are using the  inner product
notation. For functions $a(x)$ and $b(x)$, the inner product is represented as
$\big<a,b\big>=\int a(x)b(x)dx$.

\subsection{Labeled RFS and LMO Density}
The labeled single-object state $\bx$ is constructed by augmenting
a state $x\in\mathbb{X}$ with a label $\ell\in\mathbb{L}$. The labels are usually
drawn from a discrete label space, $\mathbb{L}=\{\alpha_i,i\in\mathbb{N}\}$, where
all $\alpha_i$ are distinct and the index space $\mathbb{N}$ is the set of positive
integers. A \textit{labeled RFS} is an RFS whose elements are identified by distinct labels \cite{refr:label_1,refr:label_2}. A \textit{labeled RFS} with (kinematic) state space $\mathbb{X}$ and (discrete) label space
 $\mathbb{L}$ is an RFS on $\mathbb{X}\times\mathbb{L}$ such that each realization $\mathbf{X}$ has distinct labels, i.e., $|\mathcal{L}(\mathbf{X})|=|\mathbf{X}|$. A labeled RFS and its unlabeled
 version have the same cardinality distribution. For an arbitrary labeled RFS, its multi-object density can be represented as the expression given in Lemma 1
  \cite{refr:label_6}, and our main results in this paper follow from this
 expression.
 \begin{Lem} Given an LMO density $\bpi$ on $\mathcal{F}(\mathbb{X}\times\mathbb{L})$, and for any positive integer $n$, we define the joint existence probability of the label set $\{\ell_1,\ell_2,\cdots,\ell_n\}$ by
\begin{equation}\label{joint existence probability}
  \omega(\{\ell_1,\cdots,\ell_n\})\!=\!\int\!\! \bpi(\{(x_1,\ell_1),\!\cdots,\!(x_n,\ell_n)\})d(x_1,\!\cdots\!,x_n)
\end{equation}
and the joint probability density on $\mathbb{X}^n$ of the states $x_1,\cdots,x_n$ conditional on their corresponding labels $\ell_1,\cdots,\ell_n$ by
\begin{equation}\label{joint probability density}
  P(\{(x_1,\ell_1),\cdots,(x_n,\ell_n)\})=\frac{\bpi(\{(x_1,\ell_1),\cdots,(x_n,\ell_n)\})}{\omega(\{\ell_n,\cdots,\ell_n\})}
\end{equation}
Thus, the LMO density can be expressed as
\begin{equation}\label{factorized}
  \bpi(\mathbf{X})=\omega(\mathcal{L}(\mathbf{X}))P(\mathbf{X}).
\end{equation}
\end{Lem}
\subsection{GLMB RFS Family and Its Subclasses}
\subsubsection{GLMB RFS}
GLMB RFS family  \cite{refr:label_1} is a class of tractable labeled RFS whose density is conjugate with standard multi-object likelihood function, and is closed under the multi-object Chapman-Kolmogorov equation with respect
to the standard multi-object motion model.

A GLMB RFS is a labeled RFS with state space $\mathbb{X}$ and (discrete) label space $\mathbb{L}$  distributed according to
\begin{align}\label{GLMB}
\begin{split}
\bpi_{\text{GLMB}}(\bX)=\triangle(\bX)\sum_{c\in\mathbb{C}}w^{(c)}(\mathcal{L}(\bX))[p^{(c)}]^\bX
\end{split}
\end{align}
where $\mathbb{C}$ is a discrete index set, $\Delta(\bX)$  is the distinct label indicator,
\begin{equation}
\Delta(\bX)=\delta_{|\mathcal{L}(\bX)|}(|\bX|)
\end{equation}
and $w^{(c)}(L)$, $p^{(c)}$ satisfy
\begin{align}
\begin{split}
\sum_{L\subseteq\mathbb{L}}\sum_{c\in\mathbb{C}}w^{(c)}(L)&=1\\
\int p^{(c)}(x,\ell)dx&=1.
\end{split}
\end{align}
Each term in the mixture (\ref{GLMB}) consists of a weight $\omega^{c}(\mathcal{L}(\mathcal{\bX}))$ that only depends on the labels of the multi-object state, and a multi-object exponential ${[p^{(c)}]}^\bX$
that depends on the entire multi-object state.
\subsubsection{$\delta$-GLMB RFS}
A $\delta$-GLMB RFS with state space $\mathbb{X}$ and discrete label space $\mathbb{L}$ is a special case of GLMB RFS with
\begin{equation}\label{delta-GLMB}
\begin{split}
\mathbb{C}=&\mathcal{F}(\mathbb{L})\times\Xi\\
\omega^{(c)}(L)=&\omega^{(I,\xi)}(L)=\omega^{(I,\xi)}\delta_I(L)\\
p^{(c)}=&p^{(I,\xi)}=p^{(\xi)}
\end{split}
\end{equation}
where $\Xi$ is a discrete space, i.e., it is distributed according to
\begin{equation}\label{delta-GLMB}
\bpi_{\text{$\delta$GLMB}}(\bX)=\Delta(\bX)\sum_{(I,\xi)\in\mathcal{F}(\mathbb{L})\times\Xi}\omega^{(I,\xi)}\delta_{I}(\mathcal{L}(\bX)){[p^{(\xi)}]}^\bX.
\end{equation}
\subsubsection{M$\delta$-GLMB}
An  M$\delta$-GLMB density $\bpi_{\text{M $\delta$ GLMB}}$ corresponding to the $\delta$-GLMB density $\bpi_{\text{$\delta$ GLMB}}$ in (\ref{delta-GLMB}) is a probability
density of the form
\begin{equation}\label{Mdelta-GLMB}
\bpi_{\text{M$\delta$GLMB}}(\bX)=\Delta(\bX)\sum_{I\in\mathcal{F}(\mathbb{L})}\omega^{(I)}\delta_{I}(\mathcal{L}(\bX)){[p^{(I)}]}^\bX
\end{equation}
where
\begin{equation}
\begin{split}
\omega^{(I)}=&\sum_{\xi\in\Xi}\omega^{(I,\xi)}\\
p^{(I)}(\bx,\ell)=&1_I(\ell)\frac{1}{\omega^{(I)}}\sum_{\xi\in\Xi}\omega^{(I,\xi)}p^{(\xi)}(\bx,\ell).
\end{split}
\end{equation}
\subsubsection{LMB RFS}
An LMB RFS  with state space $\mathbb{X}$, label space $\mathbb{L}$ and (finite) parameter set $\{(r^{(\ell)},p^{(\ell)}(x)):\ell\in\mathbb{L}\}$, is distributed according to
\begin{align}\label{LMB}
\begin{split}
\bpi_{\text{LMB}}(\mathbf{X})=\Delta(\bX)w(\mathcal{L}(\bX))p^\bX
\end{split}
\end{align}
where
\begin{align}
\begin{split}
\omega(I)=&\prod_{i\in\mathbb{L}}(1-r^{(i)})\prod_{\ell\in L }\frac{1_{\mathbb{L}}(\ell)r^{(\ell)}}{1-r^{(\ell)}}\\
p(x,\ell)=&p^{(\ell)}(x)
\end{split}
\end{align}
and $r^{(\ell)}$ represents the existence probability, and $p^{(\ell)}(x)$ is the probability density of the kinematic state of track $\ell$ given its existence.


\subsubsection{Labeled I.I.D Clustering RFS}
A labeled i.i.d (LIID)  clustering RFS with state space $\mathbb{X}$, label space $\mathbb{L}$ is distributed as \cite{refr:label_1}
\begin{equation}\label{LIID}
\bpi_{\text{IID}}(\bX)=\Delta(\bX)\delta_{\mathbb{L}(|\mathcal{L}(\bX)|)}(\mathcal{L}(\bX))\rho(|\mathcal{L}(\bX)|)\prod_{(x,\ell)\in\bX}\frac{v(x)}{\left<v,1\right>}
\end{equation}
where $\rho(n)$ is the cardinality distribution, $v(x)$ is the intensity of the unlabeled version of $\bpi$ and $\mathbb{L}(n)=\{\alpha_i\in\mathbb{L}\}_{i=1}^n$.
\subsubsection{Labeled Poisson RFS}
A labeled Poisson (LP) RFS   with state space $\mathbb{X}$, label space $\mathbb{L}$ is distributed as (\ref{LIID})
\begin{align}\label{LP}
\begin{split}
\bpi_{\text{Pois}}(\bX)=\Delta(\bX)&\delta_{\mathbb{L}(|\mathcal{L}(\bX)|)}(\mathcal{L}(\bX))\cdot\\&\mbox{Pois}_{\left<v,1\right>}(|\mathcal{L}(\bX)|)\prod_{(x,\ell)\in\bX}\frac{v(x)}{\left<v,1\right>}
\end{split}
\end{align}
where   $\mbox{Pois}_{\lambda}(n)=\exp^{-\lambda}\lambda^n\big/n!$  is a Poisson distribution with rate $\lambda$, $v(x)$ is the unlabeled version of $\bpi_{\text{Pois}}$ and $\mathbb{L}(n)=\{\alpha_i\in\mathbb{L}\}_{i=1}^n$.

\subsection{$\delta$-GLMB Density Approximation of LMO Density}
An  arbitrary LMO density can be approximated as a tractable $\delta$-GLMB density as shown in Lemma 2, which is applied to $\delta$-GLMB filter for generic observation model \cite{refr:label_6}.
\begin{Lem}
Given any LMO density $\bpi$ of the form (\ref{factorized}), the $\delta$-GLMB density which preserves the cardinality distribution and probability hypothesis density (PHD) of $\bpi$, and minimizes the Kullback-Leibler divergence from $\bpi$, is given by
\begin{equation}\label{GLMB approximation}
  \hat{\bpi}_{\text{$\delta$GLMB}}(\bX)=\Delta(\bX)\sum_{I\in\mathcal{F}(\mathbb{L})}\hat{\omega}^{(I)}\delta_{I}(\mathcal{L}(\bX)){[\hat{p}^{(I)}]}^\bX
\end{equation}
where
\begin{equation}\label{GLMB_where}
\begin{split}
\hat\omega^{(I)}&=\omega(I)\\
\hat{p}^{(I)}(x,\ell)&=1_I(\ell)p_{I-\{\ell\}}(x,\ell)\\
p_{\{\ell_1,\cdots,\ell_n\}}(x,\ell)&=\\
\int P(\{(x,\ell),&(x_1,\ell_1),\cdots,(x_n,\ell_n)\})d(x_1,\cdots,x_n).
\end{split}
\end{equation}
\end{Lem}
\subsection{Kullback-Leibler divergence}
In probability theory and information theory, the Kullback-Leibler divergence (KLD) \cite{refr:KLD_1, refr:KLD} is a measure of the difference between two probability distributions, and its extension to multi-object densities
$f(X)$ and $g(X)$ is given in \cite{refr:KLD_2} by
\begin{equation}\label{KLD}
D_{\text{KL}}(f;g)=\int f(X)\log\frac{f(X)}{g(X)}\delta X
\end{equation}
where the integral in (\ref{KLD}) is a set integral.
\subsection{Exponential Family}
We involve some basic concepts and propositions of information geometry. Information geometry is a theory which expresses the concepts of statistical inference in the vocabulary of differential geometry \cite{Imformation-geometry}. The ``space'' in information geometry is a space of probability distributions where each ``point'' is a distribution.
\begin{Lem}\label{exponential-family}
An exponential family has the characteristic property that it contains the exponential segment between any two of its members. The exponential segment between two distributions with densities $p(X)$ and $q(X)$ (with respect to some common dominating measure) is the one-dimensional set of distributions with densities
\begin{equation}
p_\alpha(X)={p(X)}^{1-\alpha}{q(X)}^{\alpha} e^{-\bpsi(\alpha)},\,\,\,\,0\leqslant\alpha\leqslant 1
\end{equation}
where $\bpsi(\alpha)$ is a normalizing factor such that $p_\alpha$ sums to 1.
\end{Lem}
\begin{Def}
For an exponential family $\mathcal{E}$  and for any distribution $P(\cdot)$, if the KLD from $P(\cdot)$ to any distribution $\overline Q(\cdot)$ of $\mathcal{E}$ is minimized by  $\overline Q(\cdot)=Q(\cdot)$, then $Q(\cdot)$ is  called as  ``orthogonal projection'' of $P(\cdot)$  onto $\mathcal{E}$.
\end{Def}
\begin{Lem}
The divergence $D_{\text{KL}}(P;R)$ from $P(\cdot)$ to any other distribution $R(\cdot)$ of an exponential manifold $\mathcal{E}$ can be decomposed into
\begin{equation}
D_{\text{KL}}(P;R)=D_{\text{KL}}(P;Q)+D_{\text{KL}}(Q;R)
\end{equation}
where  $Q(\cdot)$ is   ``orthogonal projection'' of $P(\cdot)$  onto $\mathcal{E}$.
\end{Lem}
\section{ Approximations for LMO Density}
In this section, we derive  three  approximations of the universal LMO density with the key statistical properties preserved or information divergence minimized, motivated by their great demands  in realworld applications,  such as some existing popular multi-object filters \cite{refr:PHD,refr:CPHD, refr:label_1, refr:label_3,refr:label_4}, the design of  new estimation algorithms or fusion algorithms.  Firstly, we derive the expressions of the first order moments of LMO density, LMB density and LP/IID clustering process in Propositions 1$-$4.   Then the  LMB  density, LP process and LIID clustering process approximations are given in Propositions 5$-$7.
\subsection{PHD of Labeled RFS}
PHD is one of the most important statistical characters of multi-object density, which is also  a strategic point in the approximation of LMO density.  A well-known filter, PHD filter\cite{refr:Mahler_book}, is designed precisely by the utilization of  PHD. 

Note that the labeled PHD $v(x,\ell)$, namely,  the PHD of the original LMO density and the unlabeled PHD $v(x)$, namely, the PHD of the unlabeled version of  the original LMO density, are related by \cite{refr:label_6}
\begin{equation}
v(x)=\sum_{\ell\in\mathbb{L}} v(x,\ell).
\end{equation}
\begin{Pro}\label{PHD-LMO}
Given an arbitrary LMO density $\bpi(\cdot)=\omega(\cdot)P(\cdot)$ on state space $\mathbb{X}$ and label space $\mathbb{L}$, the labeled PHD of $\bpi$ is
\begin{equation}\label{labeled-PHD-LMO}
v(x,\ell)=\sum_{I\in\mathcal{F}(\mathbb{L})}1_{I}(\ell)\omega(I)p_{I-\ell}(x,\ell)
\end{equation}
and the unlabeled PHD of $\bpi$ is
\begin{equation}\label{unlabeled-PHD-LMO}
v(x)=\sum_{\ell\in\mathbb{L}}\sum_{I\in\mathcal{F}(\mathbb{L})}1_{I}(\ell)\omega(I)p_{I-\ell}(x,\ell)
\end{equation}
where
\begin{align}
\begin{split}
&p_{\{\ell_1,\cdots,\ell_n\}}(x,\ell)=\\&\int P(\{(x,\ell),(x_1,\ell_1),\cdots,(x_n,\ell_n)\}) d(x_1,\cdots,x_n).
\end{split}
\end{align}
\end{Pro}
The proof of Proposition 1 is given in the Appendix A.
\begin{Pro}\label{LPHD-LMB}
Given an LMB RFS  with parameters $\bpi=\{r^{(\ell)},p^{(\ell)}(\cdot)\}_{\ell\in\mathbb{L}}$, the labeled PHD of $\bpi$ is
\begin{equation}\label{labeled-PHD-LMB}
v(x,\ell)=r^{(\ell)}p(x,\ell)
\end{equation}
with $p(x,\ell)=p^{(\ell)}(x)$.
\end{Pro}
The proof of Proposition 2 is given in the Appendix B.
\begin{Pro}\label{LPHD-LIID}
 Given an LIID clustering process $\bpi$  of the form (\ref{LIID}), the labeled PHD of $\bpi$ is
\begin{equation}\label{labeled-PHD-LIID}
v(x,\ell)=\alpha(\ell)\frac{v(x)}{\left<v,1\right>}
\end{equation}
with
\begin{equation}
\alpha(\ell)=\sum_{I\in\mathbb{L}}1_{\mathbb{L}(|I|)}(\ell)\rho(|I|).
\end{equation}
\end{Pro}

\begin{Pro}\label{LPHD-LP}
Given an LP RFS $\bpi$ of the form  (\ref{LP}), the labeled PHD of $\bpi$ is
\begin{equation}\label{labeled-PHD-LP}
v(x,\ell)=\alpha(\ell)\frac{v(x)}{\left<v,1\right>}
\end{equation}
with
\begin{equation}
\alpha(\ell)=\sum_{I\in\mathbb{L}}1_{\mathbb{L}(|I|)}(\ell)\mbox{Pois}_{\left<v,1\right>}(|I|).
\end{equation}
\end{Pro}\

\textbf{Remark:}  Observing the construction of labeled distributions in \cite{refr:label_1}, one can find that some distributions like LMB, GLMB, can use the indexes of single-object densities as the object labels. Essentially, they use different single-object densities to distinguish different objects.  However, for LP/LIID clustering process, the density-to-object strategy is not applicative, for the densities of different objects are the same.

For LP/LIID clustering process, the object labels are given based on additional information $\mathbb{L}(n)$. Observing (\ref{LP}) and (\ref{LIID}), one can find that the density of  the LP/LIID clustering process relies on the setting of the label set $\mathbb{L}(n)$.
Also, the parameter $\alpha(\ell)$ in (\ref{labeled-PHD-LIID}) and (\ref{labeled-PHD-LP}) which means the existence of object $\ell$ also relies on $\mathbb{L}(n)$.


\subsection{The Approximations of LMO density}
The LMB density, LP process, LIID clustering process approximations are given in the following. 
\begin{Pro}\label{App-LMB}
Given an arbitrary LMO $\bpi(\cdot)=\omega(\cdot)P(\cdot)$,
the LMB  RFS which matches the labeled first order moment of $\bpi$, and minimizes the Kullback-Leibler divergence from $\bpi$,  is $\hat{\bpi}_{LMB}=\{(\hat r^{(\ell)},\hat p^{(\ell)}(\cdot))\}$, where
\begin{align}
\label{LMB_where_r} \hat r^{(\ell)}&=\sum_{I\in\mathcal{F}(\mathbb{L})}1_I(\ell)\omega(I)\\
\label{LMB_where_p} p^{(\ell)}(x)&=\frac{1}{\hat r^{(\ell)}}\sum_{I\in\mathcal{F}(\mathbb{L})}1_I(\ell)\omega(I)\hat p_{I-\{\ell\}}(x,\ell).
\end{align}
\end{Pro}
The proof of Proposition 5 is given in the Appendix C.
\begin{Pro}\label{App-LIID}
Given an arbitrary LMO $\bpi(\cdot)=\omega(\cdot)P(\cdot)$, the LIID clustering process which matches the unlabeled  first order moment  and the cardinality distribution of $\bpi$  is
\begin{align}\label{poisson}
\begin{split}
&\hat \bpi_{\text{LIID}}(\bX)\\&=\Delta(\bX)\delta_{\mathbb{L}(|\mathcal{L}(\bX)|)}(\mathcal{L}(\bX))\hat \hat{\rho}(|\mathcal{L}(\bX)|)\prod_{(x,\ell)\in\bX}\frac{\hat v(x)}{\left<\hat v,1\right>}
\end{split}
\end{align}
where
\begin{equation}\label{poisson_where}
\begin{split}
  &\hat\rho(n)=\sum_{I\in\mathcal{F}_n(\mathbb{L})}\omega(I)\\
  &\hat v(x)=\sum_{\ell\in\mathbb{L}}\sum_{I\in\mathcal{F}(\mathbb{L})}1_{I}(\ell)\omega(I)p_{I-\ell}(x,\ell).
\end{split}
\end{equation}
\end{Pro}



\begin{Pro}\label{App-LP}
Given an arbitrary LMO density $\bpi(\cdot)=\omega(\cdot)p(\cdot)$, the LP  process which matches the unlabeled  first order moment  of $\bpi$  is given by
\begin{align}\label{poisson}
\begin{split}
&\hat \bpi_{LP}(\bX)\!\\&=\!\Delta(\bX)\delta_{\mathbb{L}(|\mathcal{L}(\bX)|)}(\mathcal{L}(\bX)) \mbox{Pois}_{\left<\hat v,1\right>}(|\mathcal{L}(\bX)|)\!\!\!\!\prod_{(x,\ell)\in\bX}\!\!\frac{\hat v(x)}{\left<\hat v,1\right>}
\end{split}
\end{align}
where
\begin{equation}\label{App-LP-where}
\begin{split}
   &\hat v(x)=\sum_{\ell\in\mathbb{L}}\sum_{I\in\mathcal{F}(\mathbb{L})}1_{I}(\ell)\omega(I)p_{I-\ell}(x,\ell).
\end{split}
\end{equation}
\end{Pro}

\textbf{Remark:} As the labeled PHD of    LP/LIID clustering process relies on the additional label information $\mathbb{L}(n)$, we can only provide the results matching the unlabeled PHD for  LP/LIID clustering process.

\section{Comparisons and Discussions for the Approximations of LMO Density }
 Approximation of multi-object density plays a key role in the multi-object estimation especially for the engineering implementation of algorithms. In this section, we provide a detailed comparative analyses between GLMB density, LMB density, LP process and LIID clustering process based approximations given in Lemma 2 and Propositions 5$-$7, in terms of the approximation error and computational complexity. Also some practical suggestions for the selection of approximations are also given.  One can select the proper approximation depending on the practical requirements, such as accuracy requirement, hardware ability, etc.
 \subsection{Approximation Error}
 In this subsection, we analysis approximation error of the aforementioned approximations in terms of five important aspects: correlation between objects, cardinality distribution, KLD from original density to its approximations, the accuracy of individual object spatial distribution, and the relationship between correlation and approximation error.
 \subsubsection{Correlation between objects}
In multi-object inference, statistical correlation between objects usually comes from the ambiguous
observation (which has relationship with multiple objects) when considering
the posterior multi-object density \cite{refr:label_6,refr:label_7,refr:JMB,closely-spaced-1}, or from the interactions
between objects in Markov point processes \cite{refr:Markov_PP}.  When objects strongly depend on
each other, multi-object estimation may have a  deviation if one does
not consider the correlation \cite{refr:MeMber_filter2}; otherwise, if objects are indeed independent   of each other, the independence can be utilized to simplify the multi-object density \cite{closely-spaced-5,closely-spaced-6}.

\begin{table}[htp]
\caption{Correlation Analysis}
\begin{center}
\begin{tabular}{c|c|c|c|c|c}
\hline
\hline
Approximations& LMO & GLMB &LMB &LP &LIID \\
\hline
\tabincell{c}{Correlation between\\objects}  & $c_1$  &$c_2$&$c_3$&$c_3$&$c_3$\\
\hline
\end{tabular}
\end{center}
\end{table}
$c_1$: Complete correlation

$c_2$: Preserving a part of correlation

$c_3$: Independence
\\

Table I provides the situations of correlation for the LMO density and its different approximations. The LMO density in (\ref{factorized}) is a generalized expression  able to depict any labeled RFS with any correlation between objects. After the LMO density is approximated as a GLMB density, a part of correlation  is lost, for each joint probability densities on $\mathbb{X}$ conditional on their corresponding  labels are approximately factorized as the product of its marginals. Then after the LMO density is approximated as an LMB density, LP process, or LIID clustering process, the correlation between set objects is totally lost.
\subsubsection{Cardinality Distribution}
Cardinality distribution is a very important statistical property of RFS.  In multi-object estimation, it is used to estimate the number of objects. If  the approximated density does not preserve the original cardinality distribution, the estimate of object number has a chance to be biased \cite{refr:CPHD}. Take a instance, the performance enhancement of cardinalized PHD filter \cite{refr:CPHD} towards PHD filter \cite{refr:PHD} is obtained by preserving cardinality distribution.

For an arbitrary LMO density $\bpi(\cdot)=\omega(\cdot)p(\cdot)$, its cardinality distribution is
\begin{equation}\label{cardinality_0}
\rho(n)=\sum_{I\in\mathcal{F}_n(\mathbb{L})}\omega_{I}.
\end{equation}

In GLMB density approximation,  $\hat\omega(I)=\omega(I)$ for each $I\in\mathbb{L}$, thus the approximated cardinality distribution is the same as the original one according to (\ref{cardinality_0}).

The cardinality distribution of LMB density approximation is given by
\begin{equation}\label{cardinality-LMB}
\rho_{\text{LMB}}(n)=\prod_{j\in\mathbb{L}}\left(1-r^{(j)}\right)\sum_{I\in\mathcal{F}_n(\mathbb{L})}\prod_{\ell\in I}{\frac{r^{(\ell)}}{1-r^{(\ell)}}}.
\end{equation}
Comparing (\ref{cardinality-LMB}) with (\ref{cardinality_0}), one can easily find that the LMB density approximation does not preserve the original cardinality distribution. However, note that the mean of the approximated cardinality $\rho_{\text{LMB}}(n)$ is
\begin{equation}
\sum_{n=0}^\infty n\rho_{\text{LMB}}(n)=\sum_{\ell\in\mathbb{L}}\sum_{I\in\mathbb{L}}1_I(\ell)\omega(I)
\end{equation}
which correctly matches the mean of the original cardinality distribution.

As for the LIID clustering process  and LP process approximations, the former preserves the cardinality distribution while the latter do not.
The analytical results is summarized in Table II.

\begin{table}[htp]
\caption{The Preservation of Cardinality Distribution}
\begin{center}
\begin{tabular}{c|c|c|c|c}
\hline
\hline
Approximations& GLMB &LMB &LP &LIID \\
\hline
Preserve cardinality distribution & Yes & No &No &Yes\\
\hline
\end{tabular}
\end{center}
\end{table}

\subsubsection{KLD from  Original Density to Its approximations}
We analyze the approximation errors  in term of KLD  in the framework information geometry.

$\bullet$ \textbf {KLD  for the full density $\bpi(\cdot)$}:
It follows from  \cite{Fantacci-BT} that both the M$\delta$-GLMB and LMB density families are exponential families shown in Lemma 3. Notice that the $\delta$-GLMB density approximation $\hat \bpi_{\text{$\delta$GLMB}}$ in (\ref{GLMB approximation}) is essentially an M$\delta$-GLMB density and thus is 
the ``orthogonal projection'' of any LMO density $\bpi(\cdot)$  onto the M$\delta$-GLMB family baed on Definition 1.
Also according to Proposition \ref{App-LMB}, one can find that the LMB density approximation $\hat \bpi_{\text{LMB}}$ is the ``orthogonal projection'' of any LMO density $\bpi(\cdot)$  onto the M$\delta$-GLMB family based on Definition 1.

As LMB density family is the subclass of M$\delta$-GLMB density family,  we have that $\hat\bpi_{LMB}$ also belongs to the M$\delta$-GLMB family.  Hence, according to Lemma 4, we can obtain that
\begin{equation}\label{DKL-GLMB-LMB}
\begin{split}
D_{\text{KL}}&(\bpi;\hat \bpi_{\text{LMB}})\\&=
D_{\text{KL}}(\bpi;\hat\bpi_{\text{$\delta$GLMB}})+D_{\text{KL}}(\hat\bpi_{\text{$\delta$GLMB}};\hat \bpi_{\text{LMB}})\\
&\geqslant D_{\text{KL}}(\bpi;\hat\bpi_{\text{$\delta$GLMB}}).
\end{split}
\end{equation}
\begin{figure}[htpb]
  \centering
 \includegraphics[width=8cm]{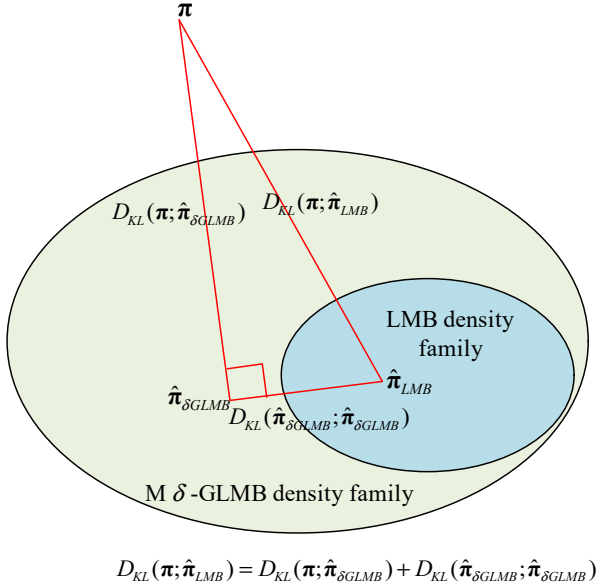}\\
  \caption{The relationship between $D_{\text{KL}}(\bpi;\hat \bpi_{\text{LMB}})$, $D_{\text{KL}}(\bpi;\hat\bpi_{\text{$\delta$GLMB}})$ and $D_{\text{KL}}(\hat\bpi_{\text{$\delta$GLMB}};\hat \bpi_{\text{LMB}})$}\label{LMO_GLMB_LMB}
\end{figure}
The relationship shown in (\ref{DKL-GLMB-LMB}) is also illustrated by Fig. 1.
Similar conclusions for LP and LIID clustering process  also can be given by,
\begin{align}
D_{\text{KL}}(\bpi;\hat \bpi_{\text{}})&\geqslant D_{\text{KL}}(\bpi;\hat\bpi_{\text{$\delta$ GLMB}})\\
D_{\text{KL}}(\bpi;\hat \bpi_{\text{LIID}})&\geqslant D_{\text{KL}}(\bpi;\hat\bpi_{\text{$\delta$ GLMB}}).
\end{align}

$\bullet$ \textbf{ {KLD for  $P(\cdot)$ and $\omega(\mathcal{\cdot})$}}:

It follows \cite{refr:computer-vision,refr:label_6} that the KLD from $\bpi$ and its approximation $\hat\bpi(\cdot)=\omega(\mathcal{L}(\cdot))P(\cdot)$  is given by
\begin{equation}\label{D_KL_App}
\begin{split}
&D_{\text{KL}}(\bpi;\hat \bpi)=C(\hat\omega)+C(\hat P)\\
\end{split}
\end{equation}
where
\begin{align}
\label{C_W}C(\hat\omega)&=D_{\text{KL}}(\omega;\overline\omega)\\
\label{C_P}C(\hat P)&=\int \log\left(\frac{P(\bX)}{\hat P(\bX)}\right)\omega(\mathcal{L}(\bX))P(\bX)\delta\bX
\end{align}
which represent the contributions to the divergence $D_{\text{KL}}(\bpi;\hat\bpi)$ from $\hat\omega(\cdot)$ and $\hat P(\cdot)$, respectively.

By substitution of $\hat \bpi_{\text{GLMB}}$, $\hat \bpi_{\text{LMB}}$, $\hat \bpi_{\text{LP}}$ and $\hat \bpi_{\text{LIID}}$ into (\ref{C_P}) and (\ref{C_W}), we can obtain the following conclusions:

\begin{itemize}
\item As for $C(\hat P)$, we have
\begin{equation}
C(\hat P_{\text{GLMB}})\geqslant C(\hat P_{\text{LMB}})\geqslant C(\hat P_{\text{LP}})\geqslant  C(\hat P_{\text{LIID}});
\end{equation}
\item $C(\hat \omega_{\text{GLMB}})$ is the lowest among all aforementioned approximations.
\end{itemize}

\subsubsection{The Accuracy of Individual object Spatial Distribution}
In the $\delta$-GLMB density approximation, any individual object $\ell$ exists in the way that it owns a unique spatial distribution for each hypothesis in which the label set $I$ includes  $\ell$,  i.e., $p^{(I)}(x,\ell)$s. Individual object may have different spatial distributions under different hypotheses due to that it may be correlated to different objects in  different ways.  The $\delta$-GLMB density approximation only discards the correlation of different objects under each hypothesis, but preserves the accuracy of individual object spatial distribution .

As for the LMB density approximation, the spatial distribution of each individual object $\ell$  in (\ref{LMB_where_p}) is extracted by weighted summation of its marginals $p_{I-\{\ell\}}(x,\ell)$ of all the  conditional joint probability densities $P(\bX)$ with $\mathcal{L}(\bX)\ni \ell$. This process also can be interpreted as a marginalization over  hypotheses  where object $\ell$ exists. During this marginalization, it loses a certain degree of  accuracy, but note that the decrease of   accuracy is slight  if this object is weakly correlated with other tracks.

Then in the LP/LIID clustering process approximations, the spatial distributions of each objects are the same and are obtained by  weighted averaging the spatial distributions of each objects (\ref{LMB_where_p}) in LMB approximation. This process also can be interpreted as a marginalization over object labels and further loses a certain degree of  accuracy. Moreover, after this marginalization, the resolution for closely spaced objects decreases,  and  target labels are lost, thus these approximations cannot enforce a one object per track constraint, which may lead to the so-called  ``spooky effect'' \cite{refr:spooky-effect-1,refr:spooky-effect-2}.
 \subsubsection{Correlation and Approximate Error}
In classical probability theory, a measure of statistical  independence is the KLD from any distribution to its ``best product'' approximation \cite{refr:KLD}. For the  LMO density in the notion of point process,
 the ``best'' LMB density approximation (given in Proposition \ref{App-LMB}) plays the same role as the  ``best product'' approximation in the Euclidean notion of density.

Any labeled RFS $\bPsi$ on space $\mathbb{X}\times\mathbb{L}$ can be represented as the union $\bPsi=\uplus_{\ell\in\mathbb{L}}\bpsi_\ell$, where each $\bpsi_{\ell}$ on space $\mathbb{X}\times \{\ell\}$ is the subset of $\bPsi$ related to object $\ell$.
Hence, $\bpsi_\ell$ is a Bernoulli RFS which is either  a singleton  $\{x,\ell\}$ with probability $r^{(\ell)}$ and density $p^{(\ell)}(x)$ or an empty set with probability $1-r^{(\ell)}$.

If $\bpsi_\ell$s are completely independent of each other, the LMO density of $\bPsi$, i.e., $\bpi(\cdot)=\omega(\mathcal{L}(\cdot))P(\cdot)$ is an LMB density and the relationships between  $\omega(\mathcal{L}(\cdot))$, $P(\cdot)$, $r^{(\ell)}$ and $p^{(\ell)}$, $\ell\in\mathbb{L}$ are the same as (\ref{LMB_where_r}) and (\ref{LMB_where_p}). Thus, the KLD from any LMO density to its ``best'' LMB approximation is a measure of statistical  independence. The larger the KLD is, the stronger correlation between objects is.

Conversely, the conclusion that the weaker correlation between objects of an LMO density is, the smaller the approximate error of ``best'' LMB approximation is, also holds.

\subsection{Computational Complexity}
We consider the computational complexity of the set integral \cite{refr:Mahler_book} of the LMO density and its approximations. In practice, the set integral usually cannot be implemented directly, but is  split in several Euclidean notion of integrals.

In GLMB density approximation, the abandon of correlation under each hypothesis makes the set integral  get rid of the curse of dimensionality,  but the combinational nature of problem still exists and the number of Euclidean notion of integrals grows exponentially with object number.

As for LMB density approximation, the marginalization over hypotheses further  solves the combinational nature of problem and the number of Euclidean notion of integrals grows linearly with object number.

Then in the LP/LIID clustering process approximation, the marginalization over object label further improves the computational efficiency and only 1 time of Euclidean notion of integral is required.

Herein, we provide the number and the dimension of Euclidean notion of integrals required in the set integral in Table III, where the  symbol $C_{M}^N$ denotes $N$ combination of $M$.
%
\begin{table}[htpb]
\caption{Computational Complexity of the Original density and Its Approximations}
\begin{center}
\begin{tabular}{c|c|c|c|c|c}
\hline
\hline
\backslashbox{$\alpha$}{$\beta$}{$\gamma$}& $\mathbb{X}$ & $\cdots$ &   $\mathbb{X}^n$ &    $\cdots$ &     $\mathbb{X}^{|\mathbb{L}|}$ \\
\hline
LMO & $C_{|\mathbb{L}|}^1$  &      $\cdots$   &  $C_{|\mathbb{L}|}^n$     & $\cdots$   & $C_{|\mathbb{L}|}^{|\mathbb{L}|}$\\
\hline
$\delta$-GLMB& $\sum_{n=1}^{|\mathbb{L}|}n\cdot C_{|\mathbb{L}|}^{n}$   &  0 &  0  &   0 & 0\\
\hline
LMB  & $|\mathbb{L}|$  &   0    &   0   &   0  &  0\\
\hline
LP/LIID & 1  &   0    &   0   &   0   &  0\\
\hline
\end{tabular}
\end{center}
\end{table}
\\
$\alpha$: Approximations
\\
$\beta$: Times
\\
 $\gamma$: The state space on which Euclidean notion of integrals perform

\subsection{Summary}
To summarize, some suggestions are provided based on the preceding analysis for the selection of different approximations in multi-object estimation,
\begin{itemize}
\item If objects are  strong correlated and the high estimate  accuracy is required, one can choose the GLMB density approximation, but note that its computational complexity is sensitive to object number.
\item If objects are weakly correlated or the requirement of  estimation accuracy is relative low, then LMB density approximation is a good choice. It provides a  good trade-off between computational burden and estimation accuracy.
\item LP/IIID clustering process approximation is also appropriate to the case where objects are weakly correlated. It has great computational efficiency, but note that it loses object identities and its resolution for closely spaced objects is low.
\end{itemize}

 \section{Numerical Results}
In this section, due to the limited space, only an simple illustration example is used to demonstrate the characteristics of different approximation methods discussed in Secion IV. Consider a labeled RFS $\bPsi$  on space $\mathbb{X}\times\mathbb{L}$, where $\mathbb{X}=\mathbb{R}$ is the field of real number and $\mathbb{L}=\{1,2,3\}$. The LMO density of $\bPsi$ is designed as
\begin{equation}\label{LMO}
\begin{split}
&\bpi(\bX)=\\
&\left\{
\begin{array}{ll}
0.01, &\bX=\emptyset\\
 0.01\mathcal{N}(x;m_1,R_1), &\bX=\{(x,1)\}\\
 0.01\mathcal{N}(x;m_2,R_2), &\bX=\{(x,2)\}\\
 0.09\mathcal{N}(x;m_3,R_3), &\bX=\{(x,3)\}\\
 0.07\mathcal{N}\left(
 \left(\begin{array}{ll}
 \!\!\!x_1\!\!\!\\
 \!\!\!x_2\!\!\!\\ \end{array}
 \right); \mathbf{m}_{12},\mathbf{R}_{12}\right), &\bX=\{(x_1,1), (x_2,2)\}\\
 0.09\mathcal{N}\left(
 \left(\begin{array}{ll}
 \!\!\!x_1\!\!\!\\
 \!\!\!x_2\!\!\!\\ \end{array}
 \right);\mathbf{m}_{13},\mathbf{R}_{13}\right),&\bX=\{(x_1,1), (x_2,3)\}\\
 0.09\mathcal{N}\left(
 \left(\begin{array}{ll}
 \!\!\!x_1\!\!\!\\
 \!\!\!x_2\!\!\!\\ \end{array}
 \right);\mathbf{m}_{23},\mathbf{R}_{23}\right),&\bX=\{(x_1,2), (x_2,3)\}\\
 0.63\mathcal{N}\left(
 \left(\begin{array}{ll}
 \!\!\!x_1\!\!\!\\
 \!\!\!x_2\!\!\!\\
 \!\!\!x_3\!\!\!\end{array}
 \right);\mathbf{m}_{123},\mathbf{R}_{123}\right), &\bX=\begin{array}{ll}\!\!\left\{(x_1,1),(x_2,2),\right.\\
 \!\!\!\left. (x_3,3)\right\}\end{array}\\
\end{array}\right.
\end{split}
\end{equation}
where
\begin{equation}
\begin{split}
&m_1=1, R_1=1\\
&m_2=2, R_2=2\\
&m_3=8, R_3=3\\
&\mathbf{m}_{12}= \left(\begin{array}{ll}
 \!\!\!1.1\!\!\!\\
 \!\!\!2.1\!\!\!\\ \end{array}
 \right), \mathbf{R}_{12}=\left[\begin{array}{cc}
1.2 &1\\
1 &2.2
 \end{array}\right]\\
 &\mathbf{m}_{13}= \left(\begin{array}{ll}
 \!\!\!1.1\!\!\!\\
 \!\!\!8.1\!\!\!\\ \end{array}
 \right), \mathbf{R}_{13}=\left[\begin{array}{cc}
1.1 &1\\
1 &1.2
 \end{array}\right]\\
 &\mathbf{m}_{23}= \left(\begin{array}{ll}
 \!\!\!2.2\!\!\!\\
 \!\!\!8.1\!\!\!\\ \end{array}
 \right), \mathbf{R}_{23}=\left[\begin{array}{cc}
2.1 &1\\
1 &1.2
 \end{array}\right]\\
  &\mathbf{m}_{123}= \left(\begin{array}{c}
 \!\!\!1.2\!\!\!\\
 \!\!\!2.2\!\!\!\\
  \!\!\!8.2\!\!\!\\ \end{array}
 \right), \mathbf{R}_{123}=\left[\begin{array}{ccc}
1.2 &2 &1\\
2& 2.2&1\\
1&1& 1.2\\
 \end{array}\right].\\
\end{split}
\end{equation}

Fig. 2 draws the decisive parameters of $\bpi$ and its approximations including $\hat \bpi_{\delta GLMB}$,  $\hat \bpi_{LMB}$, $\hat\bpi_{LP}$ and  $\hat\bpi_{LIID}$. The computing processes of different approximations are also shown in Fig. 2. It can be seen that in $\delta$-GLMB density approximation, correlation between objects is discarded under each hypothesis (different target label set $I$). Taking $I=\{1,2\}$ as instance, the contour of $p(\{(x_1,1),(x_2,2)\})$ is an oblique ellipse, while the contour of $\hat p^{(\{1,2\})}(\{(x_1,1)\hat p^{(\{1,2\})}(\{(x_2,2)\})$
 is a standard ellipse. By discarding the correlation under each hypothesis, GLMB type approximation  only needs to compute the Euclidean notion of integrals on single object space with the times 12.  In the LMB density approximation, the density of each object is obtained by marginalizing different hypotheses with $I\in\ell$. For example, following the guide-line in Fig. 2, $\hat p^{(1)}(x)$ is computed by weighted  summation of $\hat p^{(1)}(x,1)$, $\hat p^{(1,2)}(x,1)$, $\hat p^{(1,3)}(x,1)$ and $\hat p^{(1,2,3)}(x,1)$. During this marginalization, the accuracy of object spatial distribution decreases to a certain extent, but its set integral only needs to compute 3 times of Euclidean notion of integrals. In the LP/LIID clustering process approximation, following the guide-line, the spatial distributions of all objects are marginalized over the object labels and combined into one density $ {\hat v(x)}/{\left<v,1\right>}$.

In Table IV, the cardinality distribution of $\bpi$, $\hat \bpi_{\delta \text {GLMB}}$,  $\hat\bpi_{\text{LIID}}$, $\hat \bpi_{\text{LMB}}$  and $\hat\bpi_{\text{Pois}}$ are provided. Note that for the Poisson distribution $\rho_{\text{Pois}}(n)$, we only give the case $n=0,1,2,3$.  As expected, $\rho_{\delta\text{GLMB}}(n)$ and $\rho_{\text{LIID}}(n)$  are the same as $\rho(n)$, while $\rho_{\text{LMB}}(n)$ and $\rho_{\text{Pois}}(n)$ are not. Also note that the mean of  $\rho_{\text{LMB}}(n)$ and $\rho_{\text{LP}}(n)$ is 2.5 which matching the mean of $\rho(n)$.

\begin{table}[tpb]
\caption{Cardinality distribution of Original LMO Density and Its Approximations}
\begin{center}
\begin{tabular}{c|c|c|c|c}
\hline
\hline
n& 0 & 1 &2 &3  \\
\hline
\tabincell{c}{$\rho(n)$}  & 0.01  &0.11&0.25&$0.63$\\
\hline
$\rho_{\text{$\delta$GLMB}}(n)$ & 0.01  &0.11&0.25&$0.63$\\
\hline
$\rho_{\text{LIID}}(n)$  & 0.01  &0.11&0.25&$0.63$\\
\hline
$\rho_{\text{LMB}}(n)$  & 0.004  &0.068&0.352&0.576\\
\hline
$\rho_{\text{Pois}}(n)$  & 0.0821  &0.2052&0.2565&$0.2138$\\
\hline
\end{tabular}
\end{center}
\end{table}

\begin{figure*}[htpb]
  \centering
 \includegraphics[width=14cm]{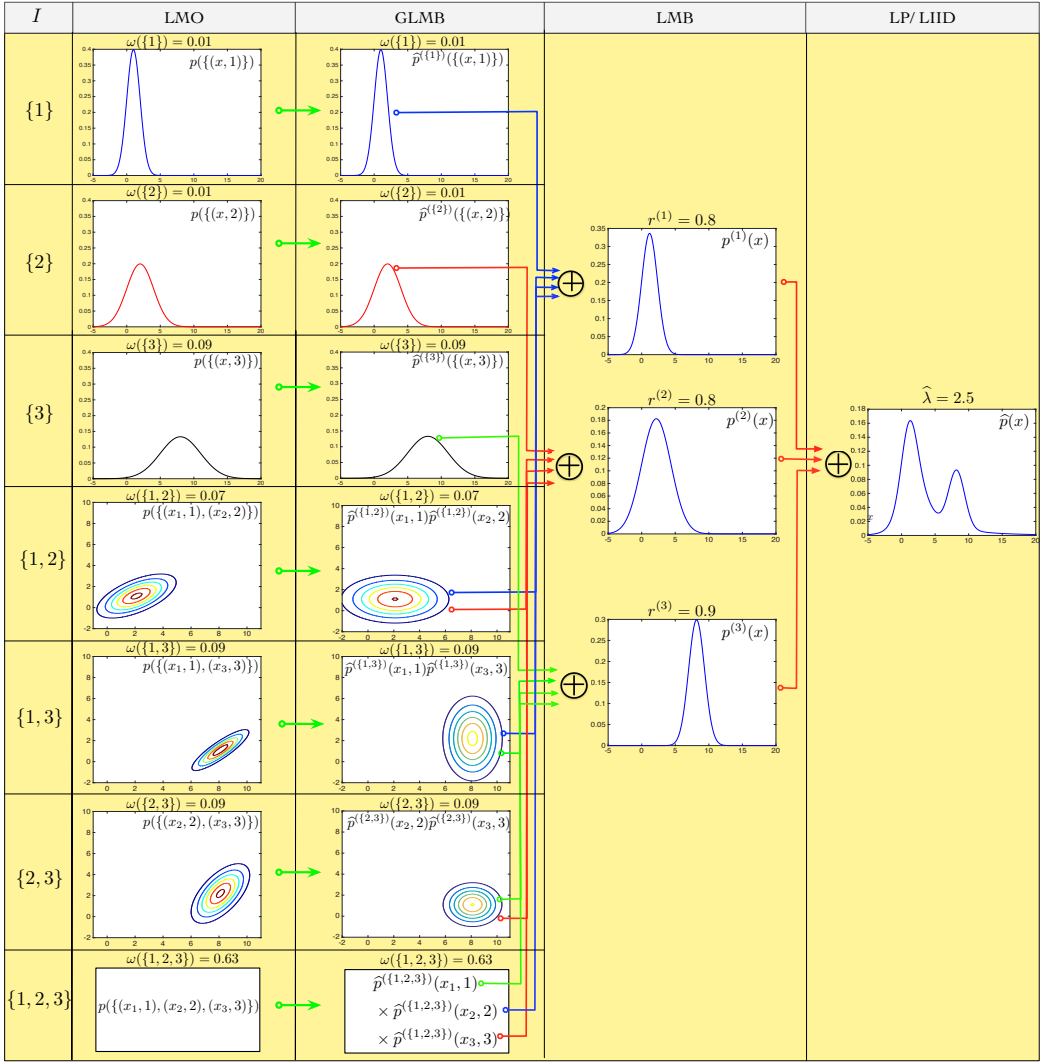}\\
  \caption{LMO density and the computing processes of its
   approximations}\label{LMO_GLMB_LMB}
\end{figure*}
\section{Conclusion}
 In this paper, we explored the approximations for the universal labeled multi-object density and derived a series of principled approximations. Moreover, we compared the relative merits of different  approximations of LMO density in terms of approximation error and computational complexity, and provided useful suggestions for  the application of these approximations based on solid theoretical analyses.  More specifically, we analyzed the approximation errors from five important aspects, namely, correlation between objects, cardinality distribution, KLD from original density to its approximations, the accuracy of individual object spatial distribution, and the relationship between correlation and approximation error.  An numerical example was also given to support our viewpoint.

\appendices
\section{Proof of Proposition \ref{PHD-LMO}}
\begin{proof}
According to the definition of PHD \cite{refr:Mahler_book}, the labeled PHD of $\bpi$ can be computed as
\begin{equation}\label{labeled-PHD-LMO-proof}
\begin{split}
  &v(x,\ell)\\
  =&\int\omega(\mathcal{L}(\{(x,\ell)\cup\mathbf{X}\}))P(\{(x,\ell)\}\cup\mathbf{X})\delta\bX\\
   =&\sum_{n=0}^\infty \frac{1}{n!}\sum_{(\ell_1,\cdots,\ell_n)\in\mathbb{L}^n}(1-1_{\{\ell_1,\cdots,\ell_n\}}(\ell))\omega(\{\ell,\ell_1,\cdots,\ell_n\})\\
   &\cdot p_{\{\ell_1,\cdots,\ell_n\}}(x,\ell)\\
   =&\sum_{n=0}^\infty \sum_{\{\ell_1,\cdots,\ell_n\}\in\mathcal{F}_n(\mathbb{L})}(1-1_{\{\ell_1,\cdots,\ell_n\}}(\ell))\omega(\{\ell,\ell_1,\cdots,\ell_n\})\\
   &\cdot p_{\{\ell_1,\cdots,\ell_n\}}(x,\ell)\\
   =& \sum_{L\in\mathcal{F}(\mathbb{L})}(1-1_{L}(\ell))\omega(\{\ell\cup L\})p_{L}(x,\ell)\\
   =& \sum_{I\in\mathcal{F}(\mathbb{L})}1_{I}(\ell)\omega(I)p_{I-\ell}(x,\ell).\\
\end{split}
\end{equation}

Moreover, based on the relationship between the labeled and unlabeled PHDs,  the unlabeled PHD is computed as
\begin{equation}\label{PHD_LMO}
v(x)=\sum_{\ell\in\mathbb{L}}v(x,\ell)=\sum_{\ell\in\mathbb{L}}\sum_{I\in\mathcal{F}(\mathbb{L})}1_{I}(\ell)\omega(I)p_{I-\ell}(x,\ell).
\end{equation}
Hence, the Proposition holds by induction.
\end{proof}

\section{Proof of Proposition \ref{LPHD-LMB}}

\begin{proof}
The LMB density also can be represented as the form of (\ref{factorized}) with
\begin{align}
\label{LMB-w}\omega(\mathcal{L}(\bX))=&\triangle(\bX)\prod_{i\in\mathbb{L}}(1-r^{(i)})\prod_{j\in \mathcal{L}(\bX) }\frac{1_{\mathbb{L}}(j)r^{(j)}}{1-r^{(j)}}\\
\label{LMB-p}P(\bX)=&p^\bX.
\end{align}
According to Proposition \ref{PHD-LMO}, substituting (\ref{LMB-w}) and (\ref{LMB-p})  into (\ref{labeled-PHD-LMO}), we can obtain the labeled PHD of $\bpi$ as,
\begin{equation}\label{labeled-PHD-LMB-proof}
\begin{split}
&v(x,\ell)\\
=&\sum_{I\in\mathcal{F}(\mathbb{L})}1_{I}(\ell)\prod_{i\in\mathbb{L}}(1-r^{(i)})\prod_{j\in I }\frac{1_{\mathbb{L}}(j)r^{(j)}}{1-r^{(j)}}p(x,\ell)\\
=&r^{(\ell)}\sum_{I\in\mathcal{F}(\mathbb{L}/\{\ell\})}\prod_{i\in I/\{\ell\}}(1-r^{(i)})\prod_{j\in I }\frac{r^{(j)}}{1-r^{(j)}}p(x,\ell)\\
=&r^{(\ell)}p(x,\ell).
\end{split}
\end{equation}
Hence, the Proposition holds.
\end{proof}
%
%
\section{Proof of Proposition \ref{App-LMB}}

\begin{proof}
Comparing  the labeled PHDs  of an arbitrary labeled RFS and the LMB RFS as shown in (\ref{labeled-PHD-LMO}) and (\ref{labeled-PHD-LMB}), we can easily obtain the parameters of the LMB density approximation matching the labeled PHD of the original LMO density $\bpi$. Specifically, the existence probability of track $\ell$ can be computed as:
\begin{equation}\label{existence_probability}
  \hat{r}^{(\ell)}=\int v(x,\ell) dx=\sum_{I\in\mathcal{F}(\mathbb{L})}1_{I}(\ell)\omega(I)
\end{equation}
and similarly the probability density of track $\ell$ conditional on its existence can be computed
\begin{equation}\label{probability_density}
  \hat{p}^{(\ell)}(x)=\frac{v(x,\ell)}{ \hat{r}^{(\ell)}}=\frac{1}{ \hat{r}^{(\ell)}}\sum_{I\in\mathcal{F}(\mathbb{L})}1_{I}(\ell)\omega(I)p_{I-\ell}(x,\ell).
\end{equation}

The KLD from $\bpi$ and any LMB RFS of the form (\ref{LMB}), is given by
\begin{equation}\label{D_KL_LMB}
\begin{split}
&D_{KL}(\bpi;\overline\bpi_{LMB})\\
=&\int \log\left(\frac{\omega(\mathcal{L}(\bX))P(\bX)}{\overline\omega(\mathcal{L}(\bX))\overline P(\bX)}\right)\omega(\mathcal{L}(\bX))p(\bX)\delta\bX\\
=&D_{KL}(\omega;\overline\omega)+\int \log\left(\frac{p(\bX)}{\overline P(\bX)}\right)\omega(\mathcal{L}(\bX))P(\bX)\delta\bX.
\end{split}
\end{equation}
First, note that
\begin{equation}\label{D_KL_p}
\begin{split}
&C(\overline P)\\
=&\int \log\left(\frac{P(\bX)}{\overline P(\bX)}\right)\omega(\mathcal{L}(\bX))P(\bX)\delta\bX\\
=&\int \omega(\mathcal{L}(\bX))P(\bX)\log P(\bX)\delta\bX-\int \omega(\mathcal{L}(\bX))P(\bX)\log \overline p^\bX\delta\bX\\
=&K_1-\int \omega(\mathcal{L}(\bX))P(\bX)\sum_{\bx\in\bX}\log \overline p(\bx) \delta\bX\\
\end{split}
\end{equation}
where $K_1$ is a constant which has no functional dependence on $\hat\bpi(\bX)$.

According to the Proposition 2a in \cite{refr:tracking-1}, i.e.,
\begin{equation}
\int \sum_{y\in Y}h(y)\pi(Y)\delta Y=\int h(y)v(y)dy
\end{equation}
with $v(\bx)$ the PHD of $\pi$, we have
\begin{equation}\label{log_p}
\int \omega(\mathcal{L}(\bX))p(\bX)\sum_{\bx\in\bX}\log\overline p(\bx)\delta\bX=\sum_{\ell\in\mathbb{L}}\int v(x,\ell)\log\overline p(x,\ell)d x
\end{equation}
with $v(x,\ell)$ shown in (\ref{labeled-PHD-LMO}).

Substituting (\ref{labeled-PHD-LMO}) and (\ref{log_p}) into (\ref{D_KL_p}), we have
\begin{equation}\label{D_KL_p2}
\begin{split}
C(\overline P)=&K_1\!\!-\!\!\sum_{\ell\in\mathbb{L}}\int\!\!\sum_{I\in\mathcal{F}(\mathbb{L})}\!\!1_{I}(\ell)\omega(I)p_{I-\ell}(x,\ell)\log\overline p(x,\ell)dx.\\
\end{split}
\end{equation}
Then (\ref{D_KL_p2}) can be rewritten as
\begin{equation}
\begin{split}
C(\overline P)=&K_1-\sum_{\ell\in\mathbb{L}}\hat r^{(\ell)}\int \hat p(x,\ell)\log\overline p(x,\ell)dx\\
=&K_1+K_2+\sum_{\ell\in\mathbb{L}}\hat r^{(\ell)}D_{KL}(\hat p(\cdot,\ell);\overline p(\cdot,\ell))
\end{split}
\end{equation}
with $K_2$ is  a constant which has no
has no functional dependence on $\hat p(x,\ell)$.

Hence,  $C(\overline P)$ is minimized only if
$\hat p(x,\ell)=p(x,\ell)$ for each $\ell\in\mathbb{L}$.

Secondly, consider the part $C(\overline\omega)=D_{KL}(\omega;\overline\omega)$.

According to the definition of KLD, we have
\begin{equation}\label{D_KL_w}
\begin{split}
&C(\overline\omega)\\
=&K_3+\sum_{I\in\mathcal{F}(\mathbb{L)}}\omega(I)\left(\sum_{\ell'\in\mathbb{L}/I}\log{\left(1-\overline r^{(\ell')}\right)}+\sum_{\ell\in I}\log{\overline r^{(\ell)}}\right)\\
=&K_3+\sum_{\ell'\in\mathbb{L}}\sum_{I\in\mathcal{F}(\mathbb{L)}}1_{\mathbb{L}/I}(\ell')\omega(I)\log\left(1-\overline r^{(\ell')}\right)+\\
&\sum_{\ell\in\mathbb{L}}\sum_{I\in\mathcal{F}(\mathbb{L)}}1_{I}(\ell)\omega(I)\log{\overline r^{(\ell)}}\\
\end{split}
\end{equation}
where $K_3$ is a constant independent of $\overline\omega(I)$.

It is obvious that
\begin{equation}
\sum_{I\in\mathcal{F}(\mathbb{L)}}1_{\mathbb{L}/I}(\ell')\omega(I)=1-\sum_{I\in\mathcal{F}(\mathbb{L)}}1_{I}(\ell')\omega(I)
\end{equation}

Thus (\ref{D_KL_w}) can be further represented as
\begin{equation}\label{D_KL_w2}
D_{KL}(\omega;\hat\omega)=K_3-\sum_{\ell\in\mathbb{L}}\left((1-\hat r^{(\ell)})\log\left({1-\overline r^{(\ell)}}\right)+\hat r^{(\ell)}\log{\overline r^{(\ell)}}\right)\\
\end{equation}
with $\hat r^{(\ell)}$ shown in (\ref{existence_probability}).

We define two Bernoulli distributions for each $\ell\in\mathbb{L}$ as
\begin{align}
&\Pr(\hat E_\ell=1)=\hat r^{(\ell)};\,\,\,\,\,\,\, \Pr(\hat E_\ell=0)=1-\hat r^{(\ell)}\\
&\Pr(\overline E_\ell=1)=\overline r^{(\ell)};\,\,\,\,\,\,\Pr(\hat E_\ell=0)=1-\overline r^{(\ell)}.
\end{align}

Then, (\ref{D_KL_w2}) yields to
\begin{equation}\label{D_KL_w3}
C(\overline\omega)=K_3+K_4+\sum_{\ell\in\mathbb{L}}D_{KL}(\Pr(\hat E_\ell=e);\Pr(\overline E_\ell=e))\\
\end{equation}
where $K_4$ is a constant having no functional dependence on each $\overline r^{(\ell)}, \ell\in\mathbb{L}$.

Hence,  $C(\overline\omega)$ is minimized only if
$\overline r^{(\ell)}=\hat r^{(\ell)}$  for each $\ell\in\mathbb{L}$.

According to (\ref{D_KL_LMB})$, D_{KL}(\bpi;\overline\bpi)$ is minimized only if both $C(\overline P)$ and $C(\overline\omega)$  are minimized. Hence, $D_{KL}(\bpi;\overline\bpi)$ is minimized by $\overline\bpi_{LMB}=\hat \bpi_{LMB}$ over the class of LMB RFS family.

%
\end{proof}
\bibliographystyle{plain}
\bibliography{refs}

\end{document}